\documentclass[reprint,notitlepage,twocolumn,
superscriptaddress,aps,longbibliography]{revtex4-1}
\usepackage[utf8]{inputenc}
\usepackage{braket}
\usepackage{amsmath}
\usepackage{amsthm}
\usepackage{mathrsfs}
\usepackage{mathtools}
\usepackage{amssymb}
\usepackage{dsfont}
\usepackage{braket}
\usepackage{bm}

\usepackage{color}
\pdfoutput=1

\usepackage[english]{babel}
\usepackage[T1]{fontenc}
\usepackage{amsmath}
\usepackage[colorlinks=true]{hyperref}
\usepackage{enumitem}
\usepackage{amsthm}
\usepackage{amssymb}
\usepackage{graphicx}
\usepackage{float}
\usepackage[ruled]{algorithm2e}

\begin{document}
\title{Controlling  Unknown Quantum States via Data-Driven State Representations}
\author{Yan Zhu}
\affiliation{QICI Quantum Information and Computation Initiative, Department of Computer Science,
The University of Hong Kong, Pokfulam Road, Hong Kong}

\author{Tailong Xiao}
\thanks{Yan Zhu and Tailong Xiao contributed equally}
\email{tailong_shaw@sjtu.edu.cn}
\affiliation{State Key Laboratory of Advanced Optical Communication Systems and Networks, Institute for Quantum Sensing and Information Processing, Shanghai Jiao Tong University, Shanghai 200240, China}%
 \affiliation{Hefei National Laboratory, Hefei, 230088, China}
\affiliation{Shanghai Research Center for Quantum Sciences, Shanghai, 201315, P.R. China}

\author{Guihua Zeng}
\affiliation{State Key Laboratory of Advanced Optical Communication Systems and Networks, Institute for Quantum Sensing and Information Processing, Shanghai Jiao Tong University, Shanghai 200240, China}%
 \affiliation{Hefei National Laboratory, Hefei, 230088, China}
\affiliation{Shanghai Research Center for Quantum Sciences, Shanghai, 201315, P.R. China}

\author{Giulio Chiribella} 
\affiliation{QICI Quantum Information and Computation Initiative, Department of Computer Science,
The University of Hong Kong, Pokfulam Road, Hong Kong}
\affiliation{Department of Computer Science, Parks Road, Oxford, OX1 3QD, United Kingdom}
\affiliation{Perimeter Institute for Theoretical Physics, Waterloo, Ontario N2L 2Y5, Canada}

\author{Ya-Dong Wu}
\email{wuyadong301@sjtu.edu.cn}
\affiliation{John Hopcroft Center for Computer Science, Shanghai Jiao Tong University, Shanghai, 200240, China}

\begin{abstract}
     Accurate control of quantum states is crucial  for quantum computing and other quantum technologies. In the basic scenario, the task is  to steer a quantum system towards a target state through  a sequence of control operations. Determining the appropriate operations, however, generally requires  information about the initial state of the system. Gathering this information becomes increasingly challenging when   the initial state is not {\em a priori} known and the system's size grows large.   
    To address this problem, we develop a machine-learning algorithm that uses a small amount of measurement data to  construct its internal representation of the system's state. The algorithm  compares this data-driven representation with a  representation of the target state, and uses reinforcement learning to output the appropriate control operations. 
We illustrate the effectiveness of the algorithm showing that it achieves accurate control of unknown many-body quantum states and non-Gaussian continuous-variable states using data from  a limited set of  quantum measurements. 
\end{abstract}

\maketitle

{\em Introduction.}
Controlling the state of noisy intermediate-scale quantum systems~\cite{preskill2018} is a major challenge for quantum  computing and other quantum technologies.  In recent years, reinforcement learning (RL)~\cite{franccois2018} has emerged as a powerful strategy for designing  quantum control policies~\cite{chen2013fidelity,PhysRevX.8.031086,PhysRevB.98.224305,zhang2019does,yao2021,borah2021,guo2021,PhysRevX.12.011059,porotti2022,metz2023self,reuer2023}. One of the benefits of this approach is that, unlike conventional model-based quantum control, RL can be used to adaptively learn quantum control policies without any knowledge of the underlying quantum dynamics~\cite{PhysRevX.12.011059}.  Beyond quantum control, RL has found applications  in quantum information science, including quantum error correction~\cite{fosel2018,nautrup2019,zeng2023}, quantum simulation~\cite{bolens2021}, quantum compilation~\cite{PhysRevLett.125.170501,fosel2021quantum,moro2021}, quantum sensing~\cite{xiao2022} and quantum communications~\cite{wallnofer2020}.

Many of the existing control methods focus on the preparation of  a known  target state from a known initial state~\cite{chen2013fidelity,PhysRevX.8.031086,zhang2019does,porotti2022,PhysRevX.12.011059,metz2023self}. In this task,  RL algorithms are often used to  learn control policies from exact state descriptions.  In practice, however, limited device calibration and imperfections in the setup can lead to uncertainty on the initial state of the system. It is then important to supplement  RL algorithms with a state characterization step, in which useful information is  gathered from  quantum measurements.  In many of the existing protocols, the relevant piece of information is the fidelity between the system's state and the target state~\cite{ chen2013fidelity,PhysRevX.8.031086,PhysRevB.98.224305,zhang2019does,porotti2022,metz2023self}.   Estimating the fidelity through measurements, however, becomes challenging as the system size grows  ~\cite{steven2011,silva2011}. To circumvent this problem, a few works  explored the possibility of directly using  measurement outcomes for reward calculation~\cite{reuer2023,borah2021,PhysRevX.12.011059}, but scalability  still remains a challenge.  


In this paper, we  introduce a representation-guided reinforcement learning (RGRL) algorithm, which combines a neural network for learning unknown quantum states with an RL algorithm to steer an uncharacterized quantum state towards a target state using only measurement statistics. The core of our  algorithm is a representation network~\cite{zhu2022} that produces its internal representation of quantum states and uses it to estimate a measure of the similarity between 
the state under control and the target state~\cite{wu2023}.  The similarity estimate is then used  
     in the RL stage to compute rewards and to determine the control operations needed to steer the quantum system to the desired state.  Overall, our algorithm provides an illustration of how a machine without any built-in knowledge of quantum physics can  learn  to control a piece of quantum  hardware.   The effectiveness of this approach is illustrated through  numerical simulations, showing strong performance for various types of many-body ground states and continuous-variable non-Gaussian states.

\begin{figure}
    \centering
    \includegraphics[width=0.45\textwidth]{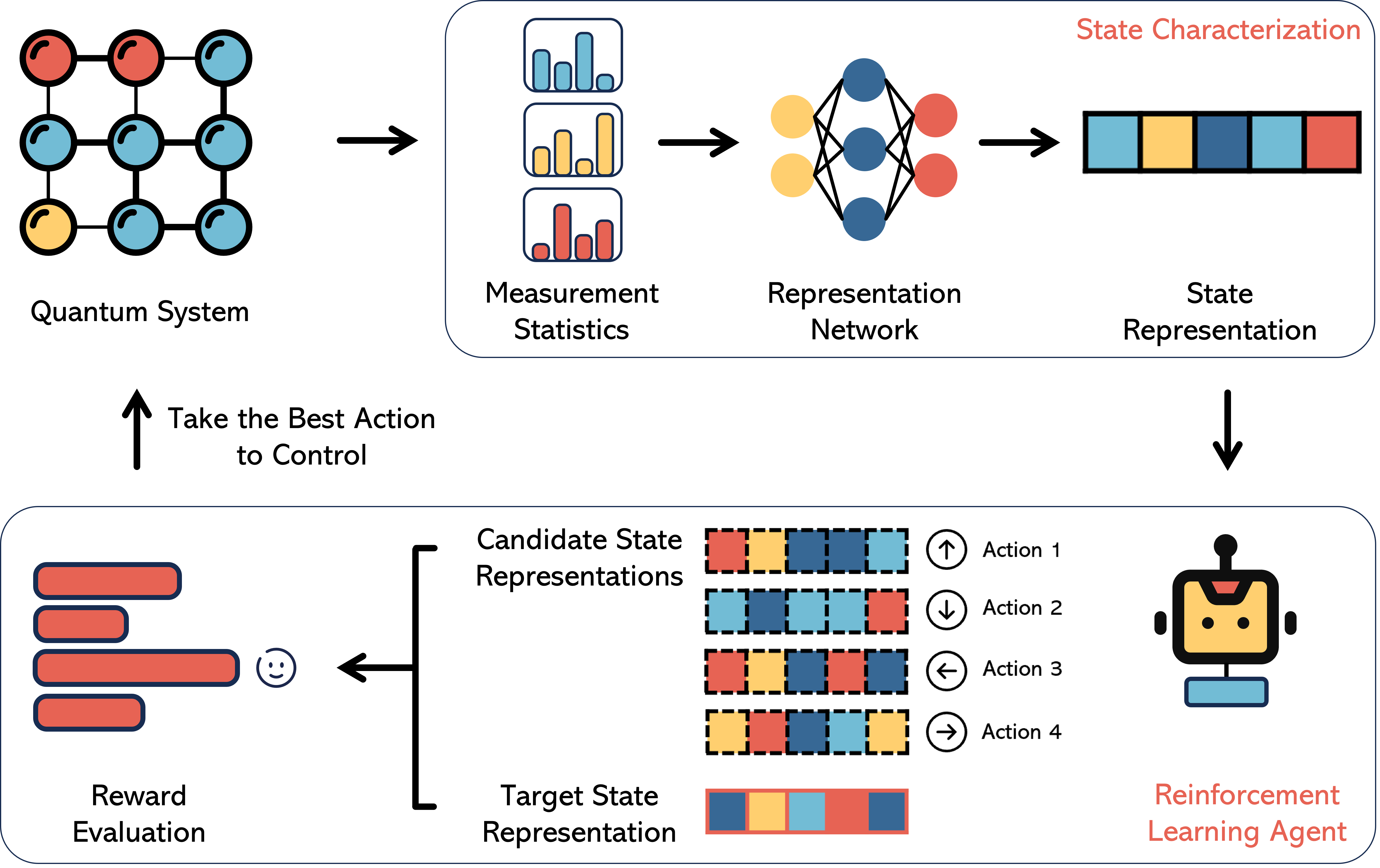}
    \caption{Schematic of our algorithm  for quantum state characterization and control.
The measurement statistics obtained by probing  a quantum system are fed into a representation network, which builds its own  representation of the quantum state. Using the state representation as input, a reinforcement learning agent evaluates the candidate state representations after performing possible control actions. The agent then selects the optimal action to maximize the expected reward, aiming to minimize the distance between the controlled state representation and the target state representation. }
    \label{fig:framework}
\end{figure}

{\em The RGRL algorithm.} Our algorithm applies to  the scenario where a classical learning agent gathers information from and performs actions upon a quantum system.   Following  RL terminology, we will refer to the quantum system as the agent's environment. The interactions between the agent and the environment are mediated by an experimenter, who performs quantum measurements and other quantum operations based on the agent's instructions.    Our  algorithm consists of  two stages: quantum state characterization/learning and quantum control, as depicted in Fig.~\ref{fig:framework}.
In the state learning stage, the agent has access to many copies of an unknown quantum state.
The set of quantum measurements that can be performed on the state is denoted by $\mathcal{M}$. In the following, we will use the notation $\bm{M}  =  (M_i)_{i=1}^n$ to denote a quantum measurement, described by an $n$-outcome positive operator-valued measure (POVM) that associates the measurement outcome $i$ to a positive operator $M_i$,  satisfying the normalization condition $\sum_{i=1}^n  M_i  =  I$.  At the beginning of the protocol,  the experimenter randomly chooses a subset $\mathcal{S} \subset \mathcal{M}$ of quantum measurements. At every control step, the experimenter performs each measurement $\bm{M} \in \mathcal{S}$ on multiple copies of the  state $\rho$, obtaining the outcome frequency distribution $\bm{d}_{M} := \text{Tr}(\rho \bm{M})$. The set of measurement statistics $\mathcal{D} := \{\bm{d}_{M}\}_{\bm{M} \in \mathcal{S}}$ is fed into a (pre-trained) representation network to yield the data-driven state representation $\bm{r}_{\rho}$. The distance between $\bm{r}_{\rho}$ and $\bm{r}_{\rho_{\text{t}}}$ in the representation space is taken as an indicator  of the distance between $\rho$ and $\rho_{\text{t}}$, where $\bm{r}_{\rho_{\text{t}}}$, i.e., the representation of the target state $\rho_{\text{t}}$, can be obtained from the simulated measurement data of $\rho_{\text{t}}$ corresponding to the measurement set $\mathcal{S}$.

At $k$th control step, the agent, associated to the RGRL algorithm, receives the  measurement data $\mathcal{D}_k$ from the environment as its partial observation and selects an action $\mathcal{A}_k$ from a finite set of predefined actions based on the policy $\pi$ and the measurement observations $\mathcal{D}_k$. Each action corresponds to a tuning of physical parameters. Upon receiving the action determined by the RGRL algorithm, the experimenter implements it on the system, driving the current quantum state to a new state in the next control step. This process is repeated in each environment-agent interaction cycle. 

Here, the control policy is modeled by a conditional probability $\pi_{\bm{\theta}}(\mathcal{A}|\mathcal{D})$ of action $\mathcal{A}$ depending on partial observations $\mathcal D$, parameterized by a neural network with parameters $\bm{\theta}$. 
At the $k$th step, a reward $r_k\propto ||\bm{r}_{\rho_k} - \bm{r}_{\rho_\text{t}}||$ is assigned, where $\rho_k$ denotes the quantum state at the $k$th step, and $||\cdot ||$ denotes the Euclidean distance in the representation space.
From the initial state to a terminal state, one full trajectory of tuples $\left\{(\mathcal D_k, \mathcal{A}_k, r_k)\right\}_{k=1}^T$ is called an episode, where $T$ denotes the maximum length of an episode.
For each episode, a  cumulative reward $R:=-\sum_{k=1}^T \gamma^{k-1} r_k$ is assigned, where $\gamma$ denotes the discount rate. The goal of the RL algorithm is to learn a policy $\pi_{\bm{\theta}}: \mathcal{D}\rightarrow \mathcal{A}$ that maximizes the averaged cumulative reward $\bar{R}$ over multiple episodes by optimizing the parameters $\bm{\theta}$ via gradient ascent, i.e., $\Delta \bm{\theta} \sim \nabla_{\bm{\theta}} \bar{R}$~\cite{sutton1999policy,silver2014deterministic}. More details about this neural network structure can be found in the supplementary material. This model-free RL algorithm can discover the optimal policy without knowledge of the explicit model of the quantum dynamics of the environment.

\begin{figure}[htbp]
    \centering
     \includegraphics[width=0.5\textwidth]{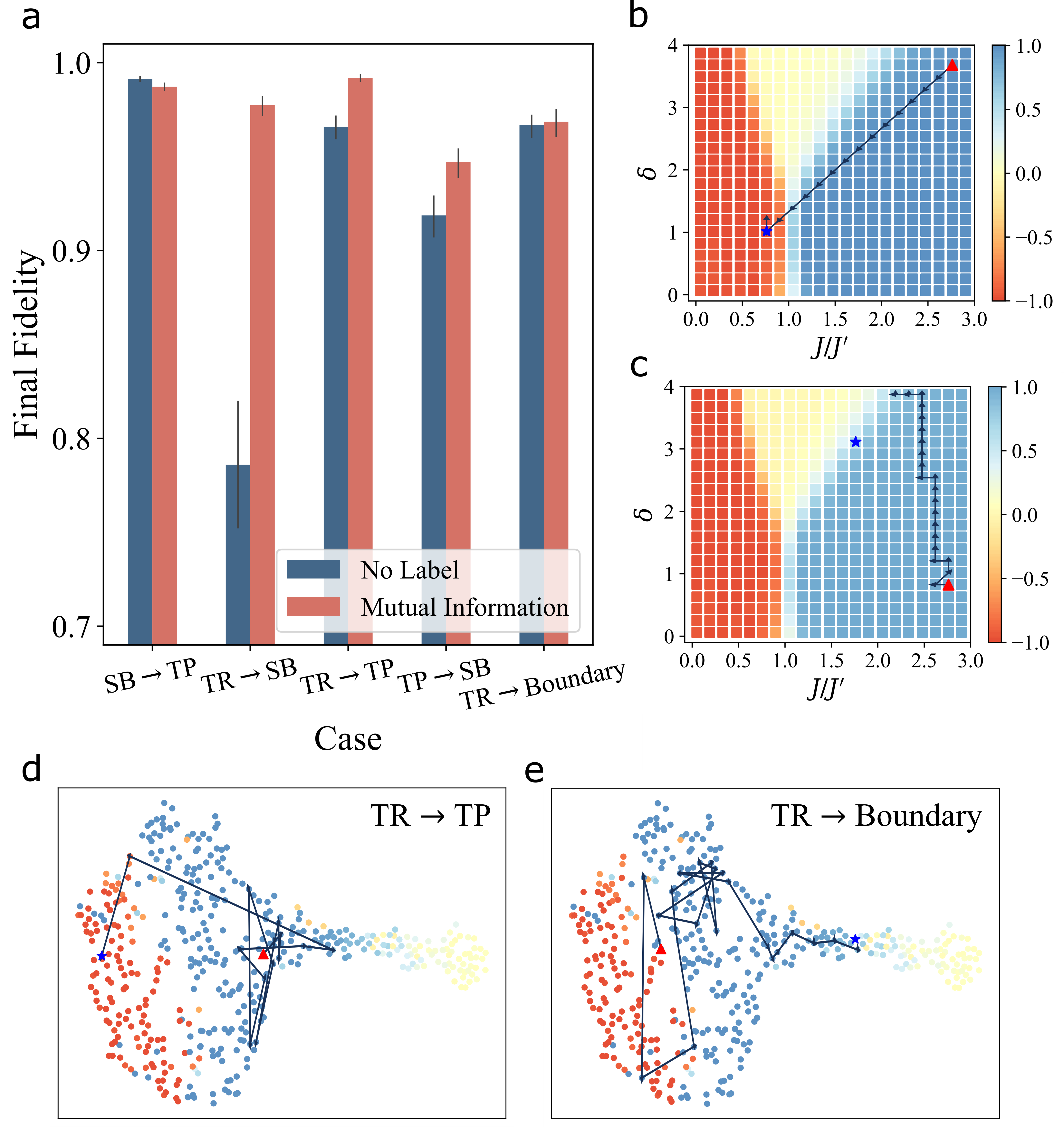}
    \caption{Control of  many-body ground states across phase transition in the bond-alternating XXZ model. Subfigure {\bf a} displays the quantum fidelities between the controlled quantum state and the target state in five different control scenarios after 30 control steps, averaged over 500 experiments, where SB denotes symmetry broken phase, TP dentoes topological SPT phase and TR denotes trivial SPT phase. The vertical lines atop each bar denote the $95\%$ confidence intervals.
Subfigures {\bf b} and {\bf c} illustrate examples of the quantum evolution trajectory of the controlled state from the trivial SPT phase to the topological SPT phase and from the trivial SPT phase to the phase boundary in the phase diagram, respectively. In these phase diagrams, the red triangle represents an initial ground state, the blue star represents a target ground state, and the red, yellow, and blue areas denote the topological SPT, symmetry broken, and trivial SPT phases, respectively.
Subfigures {\bf d} and {\bf e} depict the trajectories of the controlled state corresponding to the trajectories in Subfigures {\bf b} and {\bf c}, respectively, in the representation space, projected by the t-SNE algorithm.}
    \label{fig:XXZ}
\end{figure}

{\em Control of phase transition in many-body systems.}
 We first apply our RL algorithm for control of intermediate-scale many-body ground state across phase transition, which is of great significance in the field of many-body physics,
 particularly for those involving symmetry-protected topological phases~\cite{pollmann12,chen13,smith2022}.
 Specifically, we consider the ground states of a $50$-qubit bond-alternating XXZ model~\cite{elben2020many} respecting the reflection symmetry with respect to the bond center
\begin{align*}\notag   H_\text{XXZ}=&J\sum_{i=1}^{25}\left(\sigma_{2i-1}^x \sigma_{2i}^x +\sigma_{2i-1}^y \sigma_{2i}^y+\delta\sigma_{2i-1}^z \sigma_{2i}^z\right)  \\    &+  J'\sum_{i=1}^{24}\left(\sigma_{2i}^x \sigma_{2i+1}^x +\sigma_{2i}^y \sigma_{2i+1}^y+\delta \sigma_{2i}^z \sigma_{2i+1}^z\right). \label{eq:XXZmodel}
\end{align*}
In the parameter space $(J/J', \delta) \in (0,3) \times (0,4)$, there are three phases of matter: the topological symmetry-protected topological (SPT) phase, the trivial SPT phase, and the symmetry-broken phase, as indicated by different colors in Fig.~\ref{fig:XXZ}{\bf b}. 
We discretize this parameter space into a $21 \times 21$ grid. Assuming both the initial state and the target state fall within this grid but in different phases, we proceed as follows at each control step.

Instead of measuring every single qubit in the spin chain, we perform single-qubit Pauli measurements only on neighboring three-qubit subsystems, which correspond to a marginal of the entire system. We select $50$ different measurements, each corresponding to a Pauli string on three nearest-neighbor qubits, and use this same set of measurements at each control step. After performing the quantum measurements, we feed the measurement statistics of short-range spin correlations into the neural network. The RGRL algorithm then outputs an action to tune the pair of parameters $(J/J', \delta)$, chosen from the set of eight options: $(J/J', \delta) \leftarrow \{(J/J' \pm 0.15, \delta), (J/J', \delta \pm 0.2), (J/J' \pm 0.15, \delta \pm 0.2)\}$. 

We investigate five different control scenarios, each corresponding to a different pair of initial and target states:
(1) Control a ground state in the symmetry-broken phase towards the topological SPT phase.
(2) Control a ground state in the trivial SPT phase towards the symmetry-broken phase.
(3) Control a ground state in the trivial SPT phase towards the topological SPT phase.
(4) Control a ground state in the topological SPT phase towards the symmetry-broken phase.
(5) Control a ground state in the trivial SPT phase towards the phase boundary between the trivial SPT and symmetry-broken phases.
Figure~\ref{fig:XXZ} shows the quantum fidelity between the controlled state and the target state for these five scenarios after $30$ control steps.

To demonstrate that the quality of state representations affects the performance of quantum control, we investigate the same control scenarios using two different types of state representations. The first type is obtained by predicting the outcome statistics of measurements that have not yet been performed, as described in Ref.\cite{zhu2022}. The second type is obtained for predicting the order-two R\'{e}nyi mutual information between two subsystems of a quantum state, as outlined in Ref.~\cite{wu2023learning}. Our results in Figure.~\ref{fig:XXZ}{\bf a} indicate that using state representations designed for predicting R\'{e}nyi mutual information leads to higher quantum fidelity in our control scenarios. This improvement is attributed to the fact that these state representations capture the nonlinear properties of quantum states, thereby more clearly distinguishing different topological phases of matter.

Figure~\ref{fig:XXZ}{\bf b} shows an example of the trajectory of the ground state evolution under control, transitioning from a trivial SPT phase to a topological SPT phase. Figure~\ref{fig:XXZ}{\bf c} presents an example of a controlled ground state moving from a trivial SPT phase towards the phase boundary. Our RGRL algorithm finds the optimal path in phase space for controlling a ground state from the trivial SPT phase to the topological SPT phase, as illustrated in Figure~\ref{fig:XXZ}{\bf b}.
For controlling a ground state in the trivial SPT phase towards a state across the phase boundary, the controlled state successfully reaches the phase boundary but fails to arrive at the exact critical point. This is because the state representations obtained by our representation network are quite similar near the phase boundary, preventing the RL algorithm from accurately distinguishing different states across the boundary. As shown below, their state representations are close to each other in the representation space. More examples of ground state trajectories under control across different phases of matter are provided in the supplementary material.

Figures~\ref{fig:XXZ}{\bf d} and {\bf e} illustrate the control trajectory in state representation space, corresponding to the ground state evolution trajectories in Figures~\ref{fig:XXZ}{\bf b} and {\bf c}, projected onto a 2D plane using the t-SNE algorithm. By projecting the state representations of all ground states on the grid onto a 2D plane, we find that in representation space, the state representations do not follow their pattern in phase space. Thus, the optimal path in phase space does not correspond to the shortest trajectory in the representation space, implying that the control task we consider is highly nontrivial.

We used the state representations for predicting mutual information to plot the control trajectories in Figures~\ref{fig:XXZ}{\bf b, c, d} and {\bf e}. In the supplementary material, we also show the quantum control trajectories using the state representations for predicting measurement statistics. The results indicate that the former takes fewer control steps to arrive at the target state than the latter, demonstrating that higher-quality state representations yield higher control efficiency. Our algorithm can be easily generalized to controlling the phase transition in other similar many-body quantum systems~\cite{haldane1983,de2019}.

{\em Controlling the preparation of  continuous-variable cat states.}
Our RGRL algorithm can also be applied for controlling continuous-variable quantum states. Here, we focus on manipulating a superposition of two coherent states, known as a cat state $\ket{\mathcal C_\alpha} \propto \ket{\alpha}+\ket{-\alpha}$, where $\ket{\alpha}$ and $\ket{-\alpha}$ represent coherent states with amplitudes $\alpha\in\mathbb C$ and $-\alpha$, respectively. This cat state serves as an eigenstate of the Hamiltonian $H_\text{Kerr}:= -\hat{a}^{\dagger\, 2} \hat{a}^2 +\alpha^2 \hat{a}^{\dagger\, 2}+\alpha^{*\, 2} \hat{a}^2$~\cite{dykman2012}, with $\hat{a}$ and $\hat{a}^\dagger$ denoting annihilation and creation operators, respectively. Tuning the parameter $\alpha$ within the Hamiltonian $H_\text{Kerr}$ enables control the preparation of the cat state $\ket{\mathcal C_\alpha}$~\cite{puri2017,grimm2020}, which is useful for quantum error correction~\cite{mirrahimi2014}.

\begin{figure}[htbp]
    \centering
    \includegraphics[width=0.5\textwidth]{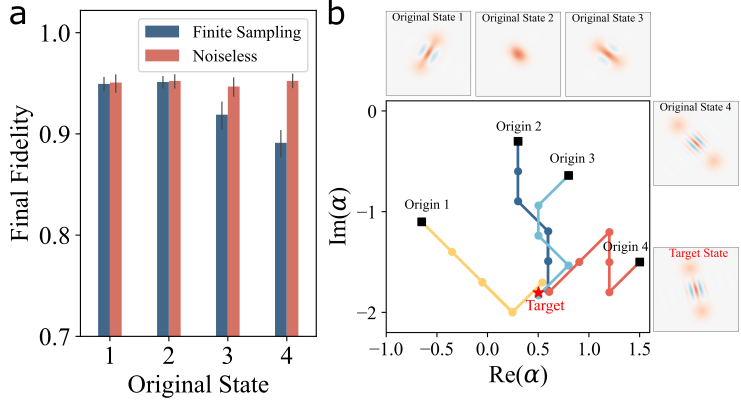}
    \caption{Control of preparing cat states. Subfigure {\bf a} shows the quantum fidelity between the controlled state and the target cat state in four different scenarios after $20$ control steps, averaged over $100$ experiments. This includes both the noiseless case and the shot-noise case. Cases 1, 2, 3, and 4 correspond to four different control scenarios depicted in Subfigure {\bf b}.  The vertical line at the top of
each bar represents the confidence interval with confidence level $95\%$.
Subfigure {\bf b} illustrates four trajectories of amplitude under quantum evolution in phase space, along with the Wigner functions of four different initial states and one target cat state. }
    \label{fig:CV}
\end{figure}

Here we consider four different control scenarios, each sharing the same target state $\ket{\mathcal C_{0.5-1.8\text{i}}^+}$, but starting from different initial states, as shown in Figure~\ref{fig:CV} (The initial state amplitudes are represented by black squares, and the target state amplitude is represented by a red star). We randomly choose three different quadratures $\left(\textrm{e}^{\textrm{i}\theta} \hat{a}^\dagger +\textrm{e}^{-\textrm{i}\theta}\hat{a}\right)/2$ associated with phases $\theta\in [0, \pi)$ and the homodyne measurements on these three quadratures form the set of measurements $\mathcal S$ at every control step. Each homodyne measurement outcome is binned into one of $100$ possibilities. At each control step, the outcome frequency distributions are fed into the neural network, and our RGRL algorithm outputs one control action out of eight options $\alpha \leftarrow \{\alpha\pm \beta, \alpha\pm \beta \text{i}, \alpha\pm \beta\pm \beta\text{i}\}$, where $\beta$ denotes the size of the parameter change at each control step.

Figure~\ref{fig:CV}{\bf a} shows the quantum fidelity between the controlled state and the target state in four different scenarios after only 20 control steps. The figure includes both cases of ideal measurement statistics and measurement statistics with shot noise due to finite sampling. In the noiseless case, the quantum fidelity always exceeds 0.95 for each initial state. In the noisy case, we simulate the shot noise by adding Gaussian white noise with a variance of 0.1 to each frequency and renormalizing the overall frequency distribution. The resulting quantum fidelity is degraded by shot noise when the initial cat state has a large amplitude.
Figure~\ref{fig:CV}{\bf b} shows the trajectories of controlled amplitudes in different control scenarios. Although the Wigner functions of the four initial states indicate significant differences, the final amplitude of the controlled state is close to that of the target state in all cases.

\begin{figure}[htbp]
    \centering
    \includegraphics[width=0.4\textwidth]{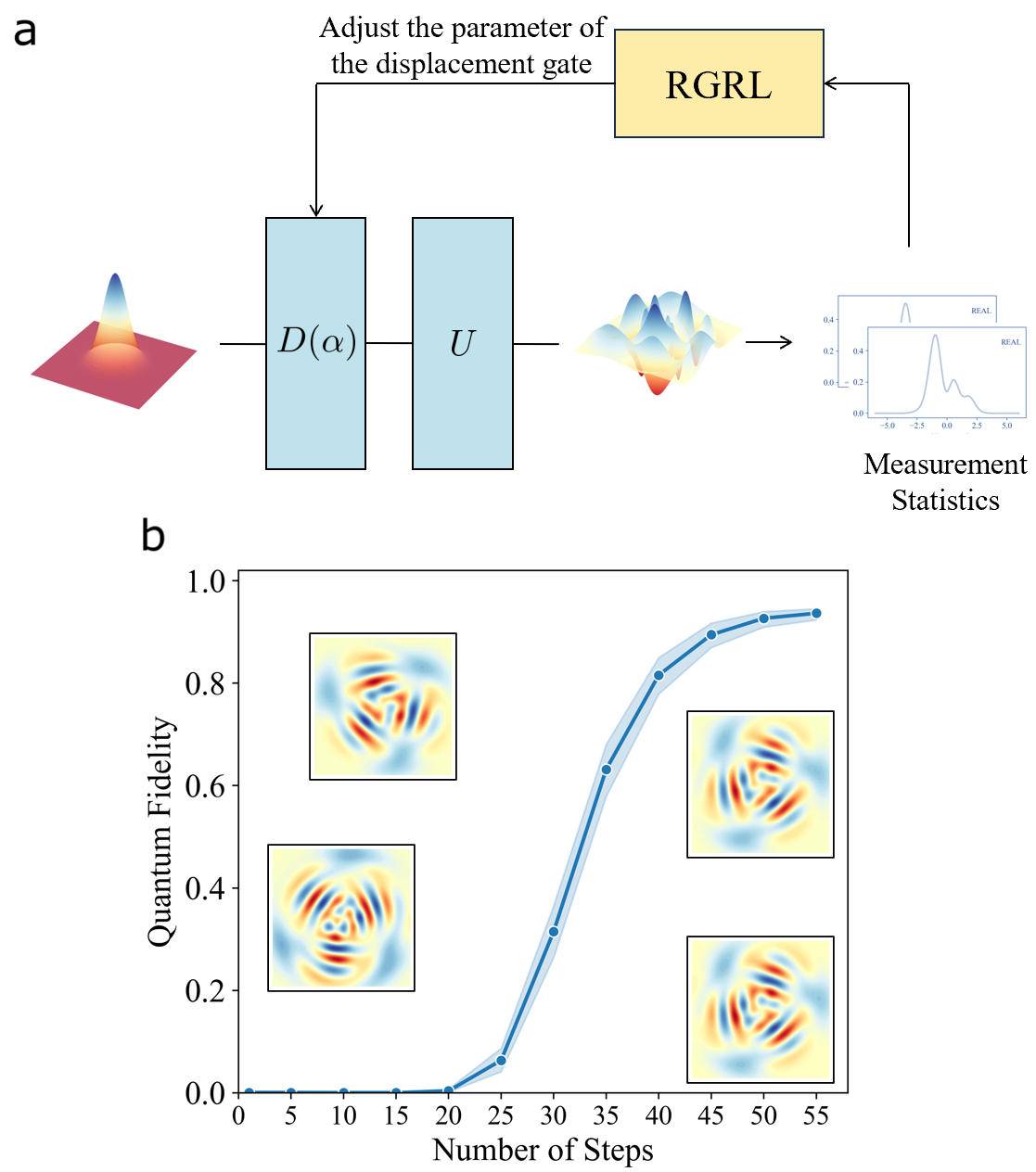}
    \caption{Producing target output of continuous-variable unitary quantum process.
Subfigure {\bf a} illustrates the measurement-and-control loop for producing the target output of a quantum process $\mathcal U$ by controlling the parameter $\alpha$ of a displacement gate applied to a vacuum state. Subfigure {\bf b} shows the quantum fidelity between the output and the target within $55$ control steps, averaged over $200$ randomly chosen pairs of initial and target states, along with several examples of Wigner function of the controlled states. }
    \label{fig:outputControl}
\end{figure}

{\em Generating the output of a non-Gaussian quantum process.}
As an additional application, we employ our RGRL algorithm to produce a target output from an unknown  quantum process.
We consider the scenario where a known target output state can be obtained by applying the unknown  quantum process to a certain input state. The goal is to find an input that causes the quantum process to approximately generate the target output.

To achieve this, we apply the unknown quantum process to multiple copies of a randomly chosen input state, resulting in multiple copies of the corresponding output state. By performing randomized measurements on the output, we estimate the similarity  between the actual  output state and the target output state. Based on the estimated similarity, we then modify  the input preparation.  This learning-and-control cycle is repeated until we identify an appropriate input that  produces an accurate approximation the target output.

As an example, we apply our algorithm to prepare target output states of a continuous-variable Kerr quantum gate. This approach can also be extended to cubic phase gates~\cite{lloyd1999,hillmann2020,yanagimoto2020}, SNAP gates~\cite{heeres2015,krastanov2015}, and other important non-Gaussian quantum gates~\cite{kudra2022}.
 The Kerr quantum gate is unitary quantum process $\mathcal U(\cdot)=\text{e}^{-\textrm{i}K }\cdot \text{e}^{\textrm{i}K }$,  where $K= \hat{a}^{\dagger\, 2}\hat{a}^2$~\cite{dykman2012}, akin to $H_{\text{kerr}}$ discussed earlier. The input states can be selected from the set of coherent states $\ket{\alpha}=D(\alpha)\ket{0}$, where $D(\alpha)$ is a displacement operator with $\alpha=r \textrm{e}^{\textrm{i}\psi}$, $r\in [0,3]$ and $\psi\in [0,2\pi)$, and $\ket{0}$ denotes a vacuum state. Three quadratures are randomly chosen as homodyne measurement bases, constituting the set of measurements $\mathcal S$ to be performed on the output at each control step. Taking the measurement statistics corresponding to the measurements in $\mathcal S$ as input, our RGRL algorithm determines the tuning of the parameter $\alpha$ in the displacement gate $D(\alpha)$ out of four options: $(|\alpha|, \text{arg}(\alpha))\leftarrow \{(|\alpha|\pm 0.09, \text{arg}(\alpha)\pm 0.06\pi)\}$.
 The measurement-and-control cycle is depicted in Fig.~\ref{fig:outputControl}{\bf a}.

Figure~\ref{fig:outputControl}{\bf b} illustrates the quantum fidelity between the generated output state and the target output state within $55$ control steps, averaged over $200$ random pairs of initial input states and target output states. The results indicate that, although the initial output state fidelity is nearly zero, the quantum fidelity exceeds 0.9 after 50 control steps. Several examples of the Wigner function of the controlled states reveal that the state during control is highly non-Gaussian, making its control particularly challenging when not fully characterized.

{\em Conclusions.}
We developed a reinforcement-learning  algorithm for steering a quantum system, initially in an unknown state, to a given target state. 
In our algorithm, each control action is determined by a neural network based on   measurement  data from a small set of quantum measurements. The measurement set is randomly chosen, independently of the target quantum state, thereby offering flexibility for a broad set of applications involving the control of uncharacterized,  intermediate-scale quantum systems using a limited set of quantum measurements. \\

{\em Acknowledgement.} YZ and GC acknowledge funding from the Hong Kong Research Grant Council through grants no.\ 17307520,  SRFS2021-7S02, and T45-406/23-R, and  the John Templeton Foundation through grant  62312, The Quantum Information Structure of Spacetime (qiss.fr). T.L. X and G.H. Z thank the support from  Shanghai Municipal Science and Technology Major Project (2019SHZDZX01) and the Innovation Program for Quantum Science and Technology (Grants No.2021ZD0300703).

\bibliography{refs}

\end{document}


\title{Supplementary Information: Controlling  Unknown Quantum States via Data-Driven State Representations}
\author{Yan Zhu}
\affiliation{QICI Quantum Information and Computation Initiative, Department of Computer Science,
The University of Hong Kong, Pokfulam Road, Hong Kong}

\author{Tailong Xiao}
\affiliation{State Key Laboratory of Advanced Optical Communication Systems and Networks, Institute for Quantum Sensing and Information Processing, Shanghai Jiao Tong University, Shanghai 200240, China}%
 \affiliation{Hefei National Laboratory, Hefei, 230088, China}
\affiliation{Shanghai Research Center for Quantum Sciences, Shanghai, 201315, P.R. China}

\author{Guihua Zeng}
\affiliation{State Key Laboratory of Advanced Optical Communication Systems and Networks, Institute for Quantum Sensing and Information Processing, Shanghai Jiao Tong University, Shanghai 200240, China}%
 \affiliation{Hefei National Laboratory, Hefei, 230088, China}
\affiliation{Shanghai Research Center for Quantum Sciences, Shanghai, 201315, P.R. China}

\author{Giulio Chiribella} 
\affiliation{QICI Quantum Information and Computation Initiative, Department of Computer Science,
The University of Hong Kong, Pokfulam Road, Hong Kong}
\affiliation{Department of Computer Science, Parks Road, Oxford, OX1 3QD, United Kingdom}
\affiliation{Perimeter Institute for Theoretical Physics, Waterloo, Ontario N2L 2Y5, Canada}

\author{Ya-Dong Wu}
\affiliation{John Hopcroft Center for Computer Science, Shanghai Jiao Tong University, Shanghai, 200240, China}

\maketitle

\section{Structure of Representation Network}

\begin{figure}[htbp]
    \centering
    \includegraphics[width=0.25\textwidth]{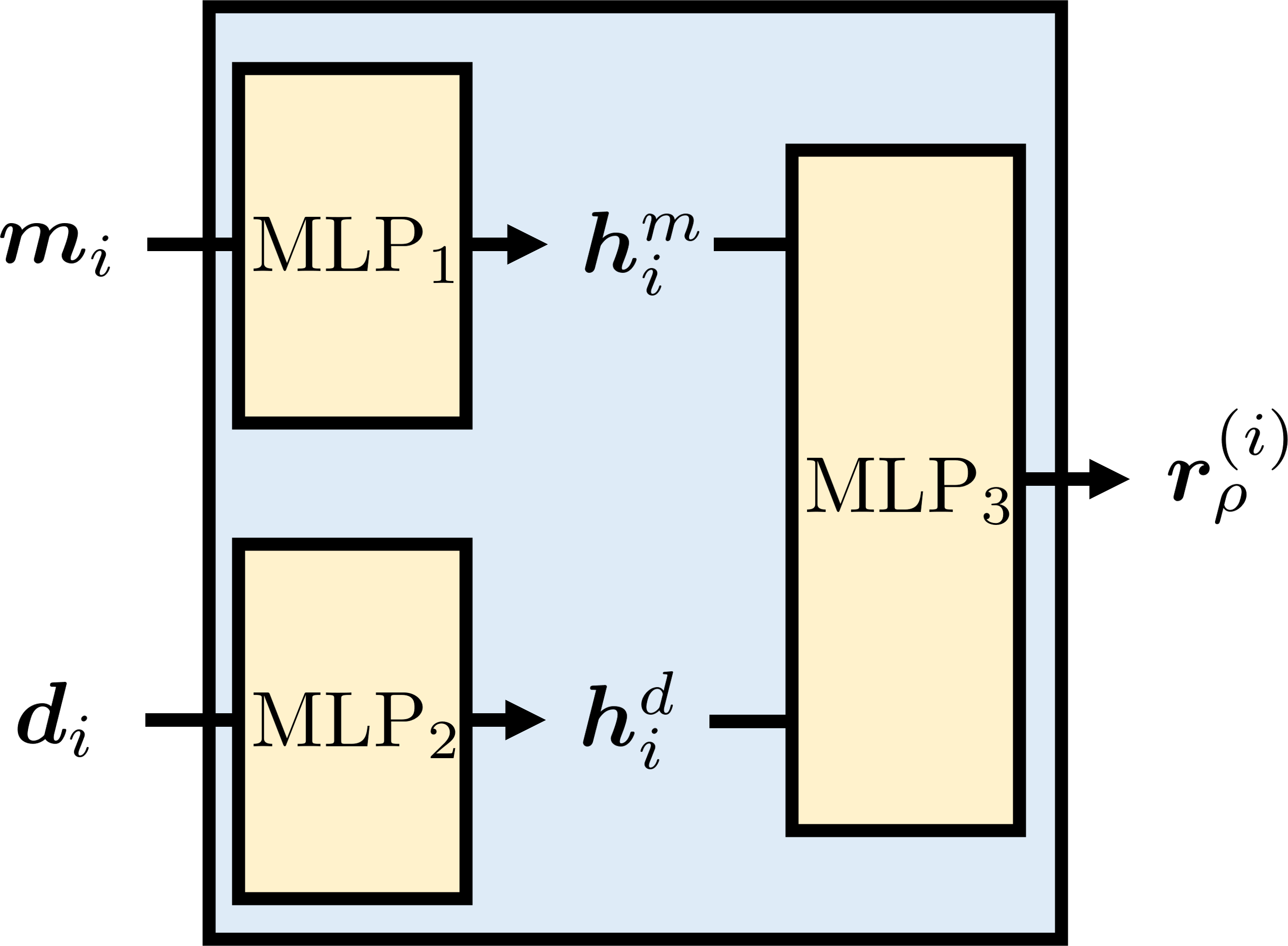}
    \caption{Structure of the representation network.}
    \label{fig:rep_struc}
\end{figure}

Here, we introduce the implementation of the representation network utilized in our framework. As illustrated in Figure~\ref{fig:rep_struc}, the representation network $ f_{\bm{\xi}} $ takes as input the parametrization $\bm{m}_i$ of measurement $M_i\in\mathcal{S}$ and its outcome statistics $\bm{d}_i$ for the specific state $\rho$. For each pair $(\bm{m}_i, \bm{d}_i)$, the representation network produces a vector $\bm{r}_i = f_{\bm{\xi}}(\bm{m}_i, \bm{d}_i)$. These vectors, corresponding to different pairs, are then combined into a single vector $\bm{r}$ by an average function.

We utilize two distinct methods to train the representation network. When only measurement data of the quantum states is available, we employ the GQNQ architecture proposed by Zhu et al. in~\cite{zhu2022}. Specifically, we conduct self-supervised learning based on this measurement data. The process involves inputting the measurement data into the representation network to generate representations of the quantum states. Subsequently, we reconstruct the measurement statistics from these representations using a generation network. During training, we minimize the reconstruction loss between the predicted measurement statistics and the ground truth.

When properties of the quantum states, such as mutual information, are available, we utilize the architecture introduced in~\cite{wu2023learning}. In this approach, the representations generated by the representation network are fed into a prediction network. During training, given the availability of labeled property data, we employ a supervised learning method to minimize the loss between the estimated and true values of these properties.

\section{Reinforcement learning algorithms}

In the main text, we introduce the basic framework of representation-guided reinforcement learning (RGRL)
algorithm. Here, we will provide more mathematical and technical details on its theory and implementation. Moreover, we also will provide detailed hyperparameter settings for the reproducibility. The environment of controlling the ground state can be viewed as a Markov decision process (MDP), therefore it is highly suited for RL \cite{sutton2018reinforcement}. We make use of model-free RL as the intelligent agent to optimize the control policies.  The observation $s$ of the RL agent is the neural representation of current ground quantum state of which belonging to a neural observation set denoted by $\mathcal{S}$. The action $a \in \mathcal{A}$ controls the adjustment of current Hamiltonian parameters. The reward is defined as the distance of current ground state and the target ground state in the representation space. More formally, the $k$th control step of the reward is given by
\begin{equation}
    r_k = -C\times \frac{\left\| \bm{r}_{\rho_k} - \bm{r}_{\rho_{\text{t}}} \right\|_2}{\sqrt{d}},
\end{equation}
where $d$ is the dimension of the neural representation, $C$ is a constant factor to scale the reward value, $\left\| \cdot \right\|_2$ denotes the Euclidean norm, $\bm{r}_{\rho_k}$, $\bm{r}_{\rho_{\text{t}}}$ are the neural representations of the $k$th state and the target state, respectively. In the following demonstration, for simplicity, we use $s$ to denote the neural representation. The goal of RL is to find an optimal policy for the agent to obtain optimal rewards. We use policy gradient methods which aims at modelling and optimizing the policy directly. The expected reward function
\begin{equation}
     J(\theta) = \sum_{s\in \mathcal{S}} d^\pi (s) \sum_{a\in \mathcal{A}} \pi_\theta(a|s) Q^\pi (s,a),
\end{equation}
where $d^\pi(s)$ is the stationary distribution of Markov chain for policy $\pi_\theta$, $Q^\pi(s,a)$ is the value function when given by a policy $\pi$. $\mathcal{A}, \mathcal{S}$ denote the action space and observation spaces, respectively.  Here, we omit notation $\theta$ when the policy $\pi_\theta$ is present in the subscript of other functions. The policy gradient theorem \cite{sutton1999policy} states that the gradient over the reward function is given by
\begin{equation}\label{policy_gradient}
\begin{split}
    \nabla_\theta J(\theta) & \propto \sum_{s\in \mathcal{S}} d^\pi (s) \sum_{a\in \mathcal{A}} Q^\pi(s,a) \nabla_\theta \pi_\theta(a|s), \\
    & = \mathbb{E}_\pi \left[ Q^\pi(s,a)\nabla_\theta \ln \pi_\theta(a|s)  \right],
\end{split}
\end{equation}
where $\mathbb{E}_\pi$ refers to $\mathbb{E}_{s\sim d^\pi, a\sim \pi_\theta}$ when both state and action distributions follow the policy $\pi_\theta$.  Eq.~(\ref{policy_gradient}) lays the foundation of policy gradient algorithms in RL. Meanwhile, Eq.~(\ref{policy_gradient}) has no bias but have large variance. Many methods focus on reducing the variance of the estimated gradient while keeping the bias unchanged. Therefore, during updating the policy parameters, we often use the advantage function $A(s,a) = Q(s,a) - V(s) $ rather than the $Q$ function. $V(s)$ is the state value function used to evaluate the expected reward of current state whatever actions it takes. To better estimate the state value function, we often use Actor-Critic architecture to model the policy gradient algorithm. Critic network is used to estimate the state value function and actor network is used to model the policies. 

In order to improve the sample efficiency and exploration ability, off-policy gradient methods are often employed. More formally, suppose the generated data trajectories obey the behavior policy $\beta(a|s)$, the objective function sums up the reward over the state distribution defined by this behavior policy
\begin{equation}
\begin{split}
    J(\theta)&=\sum_{s \in \mathcal{S}} d^\beta(s) \sum_{a \in \mathcal{A}} Q^\pi(s, a) \pi_\theta(a \mid s), \\
    &=\mathbb{E}_{s \sim d^\beta}\left[\sum_{a \in \mathcal{A}} Q^\pi(s, a) \pi_\theta(a \mid s)\right],
\end{split}
\end{equation}
where where $d^\beta$ is the stationary distribution of the behavior policy $\beta$. Note that $\pi$ refers to the target policy which is used to estimate the state-action function, i.e. $Q(s,a)$. Subsequently, we can rewrite Eq.~(\ref{policy_gradient}) as follows
\begin{equation}\label{off-gradient}
    \begin{split}
        \nabla_\theta J(\theta) =\mathbb{E}_\beta\left[\frac{\pi_\theta(a | s)}{\beta(a | s)} Q^\pi(s, a) \nabla_\theta \ln \pi_\theta(a | s)\right],
    \end{split}
\end{equation}
where we call $\frac{\pi_\theta(a | s)}{\beta(a | s)} $ as importance weight. The off-policy gradient Eq.~(\ref{off-gradient}) implies that we can use behaviour policy generated trajectories to update the policy parameters. One important fact is that  we omit the term of gradient over $Q$ function i.e. $\nabla_\theta Q^\pi(s,a)$. Fortunately, it turns out that approximated gradient with the gradient of $Q$ ignored still guarantees the policy improvement and eventually achieve the true local minimum \cite{degris2012off}. 

Trust region policy optimization (TRPO) algorithm \cite{schulman2015trust} is used to stabilize the training process, namely avoiding parameter updates that change the policy too much at one step. Consider the objection function, 
\begin{equation}\label{TRPO_J}
    J(\theta)=\sum_{s \in \mathcal{S}} \rho^{\pi_{\theta_{\text {old }}}} \sum_{a \in \mathcal{A}}\left(\pi_\theta(a | s) \hat{A}_{\theta_{\text {old }}}(s, a)\right),
\end{equation}
where $\rho^\pi$ denotes the distribution over states following the policy $\pi$,
$\theta^{\text{old}}$ is the policy parameters before the update, $\hat{A}$ denotes the estimated advantage function. Let the behaviour policy $\beta$ be $\pi_{\theta_{\text{old}}}$ and combines the advantage function, Eq.~(\ref{TRPO_J}) can be rewritten as 
\begin{equation}\label{TRPO_new}
    J(\theta) = \mathbb{E}_{s\sim \rho^{\pi_{\text{old}}}, a\sim \pi^{\text{old}}} \left[ \frac{\pi_\theta(a|s)}{\pi^{\text{old}}(a|s)} \hat{A}_{\theta_{\text{old}}}(s,a) \right ].
\end{equation}
TRPO aims to maximize Eq.~(\ref{TRPO_new}) by using policy gradient methods. Besides, TRPO constructs an extra constraint to enforce the parameter update not causing large variance, that is 
\begin{equation}\label{KL-cons}
    \mathbb{E}_{s \sim \rho^{\pi_{\theta_\text{old }}}}\left[D_{\mathrm{KL}}\left(\pi_{\theta_{\text {old }}}(. \mid s) \| \pi_\theta(. \mid s)\right] \leq \delta\right. ,
\end{equation}
where $D_{\text{KL}}$ denotes the KL divergence of two probability distributions. By constraint (\ref{KL-cons}), the old and new policy will not be large so that the training process is stabilized. However, although TRPO has beautiful theoretical formation, it is time-consuming to calculate the KL divergence in reality. Therefore, we consider using a more efficient algorithm called proximal policy optimization (PPO) \cite{schulman2017proximal} to simplify the calculation. 
PPO uses a clipped objection function and it is given by
\begin{widetext}
    \begin{equation}
    J^{\mathrm{CLIP}}(\theta)=\mathbb{E}_{s\sim \rho^{\pi_{\text{old}}}, a\sim \pi^{\text{old}}}\left[\min \left(r(\theta) \hat{A}_{\theta_{\text {old }}}(s, a), \operatorname{clip}(r(\theta), 1-\epsilon, 1+\epsilon) \hat{A}_{\theta_{\text {old }}}(s, a)\right)\right],
\end{equation}
\end{widetext}
where $r(\theta)  =\frac{\pi_\theta(a|s)}{\pi^{\text{old}}(a|s)} $ denotes the ratio between the old and target policy, the function $\text{clip}(r(\theta), 1-\epsilon, 1+\epsilon)$ clips the ratio to be no more than $1+\epsilon$ and no less than $1-\epsilon$. The objective function of PPO takes the minimum one between the original value and the clipped one. As a result,  we lose the motivation for increasing the policy update to extremes for better rewards. In implementation, to encourage the exploration, the objective function is given by
\begin{widetext}
    \begin{equation}
    J^{\mathrm{CLIP}^{\prime}}(\theta)=\mathbb{E}_{s\sim \rho^{\pi_{\text{old}}}, a\sim \pi^{\text{old}}}\left[J^{\mathrm{CLIP}}(\theta)-c_1\left(V_\theta(s)-V_{\text {target }}\right)^2+c_2 H\left(s, \pi_\theta(\cdot)\right)\right],
\end{equation}
\end{widetext}
where $c_1$ and  $c_2$ are two hyperparameter constants, $H(\cdot)$ denotes the entropy function, $V_{\text{target}}$ denote the sate value which can be calculated by using the sampled trajectories. This error term is generally added in Actor-Critic architecture. The pseudocode of PPO algorithm is presented in Algorithm \ref{PPO}. 

\begin{algorithm*}[htbp]
\SetAlgoVlined
  \caption{PPO-Clip for RGRL algorithm }\label{alg:two}
  \label{PPO}
   \SetKwInOut{Input}{input}\SetKwInOut{Output}{output}
\Input {Initial policy (actor) parameters $\theta_0$, initial value (critic) function parameters $\phi_0$; The maximum length (horizon) of each episode $T$.}
\BlankLine
\For{$k = 0,1,2,\cdots $ }{
Collect sampled trajectories $\mathcal{D}_k = \{\tau_i\}$ by running current policy $\pi_k=\pi(\theta_k)$ in the environment with $\tau_i = (\bm{r}_{\rho_i}, a_i, r_k, s_{\rho_{i+1}})$. \\
Compute the return (rewards-to-go) $\hat{G}_t$;\\
Compute advantage function estimates, $\hat{A}_t(s_t,a_t) = Q^\pi(s_t,a_t)-V_{{\phi_k}}^\pi(s_t)$ based on the current value function $V^\pi_{\phi_k}$. \\
Update the policy by maximizing the PPO-Clip surrogate, 
\[
\theta_{k+1}=\arg \max _\theta \frac{1}{\left|\mathcal{D}_k\right| T} \sum_{\tau \in \mathcal{D}_k} \sum_{t=0}^T J^{\text{CLIP}^\prime}(\theta_k),
\]
via stochastic gradient ascent with Adam method \cite{kingma2014adam}.\\

Fit value function by regression on mean-squared error:
\[
\phi_{k+1}=\arg \min _\phi \frac{1}{\left|\mathcal{D}_k\right| T} \sum_{\tau \in \mathcal{D}_k} \sum_{t=0}^T\left(V_\phi\left(s_t\right)-\hat{G}_t\right)^2,
\]
typically also via stochastic gradient descent algorithm such as Adam.\\
}
\end{algorithm*}

We note that PPO-Clip algorithm used in this work is not the only RL algorithm. Other RL algorithms such as A3C \cite{mnih2016asynchronous}, DDPG \cite{lillicrap2015continuous} and SAC \cite{haarnoja2018soft} can also be used to construct our RGRL algorithm. Note that chooing the best suited RL algorithm is not the focus of our work. In this work, the proposed model-free RL can be improved by using a model-based RL algorithm \cite{moerland2023model}, which may further reduce the sampling overhead.

\section{Hyperparameter specifications in RGRL}
The hyperparameters of PPO-Clip based RGRL algorithm are presented in detail  in this section. In general, RL algorithm has more hyerparameters than supervised machine learning models.  It generally includes the number of layers, the number of neurons of each layer with respect to actor and critic neural network,  and the learning rate $\alpha$. Except for the general hyperparameters, particular RL parameters includes the total number of steps $M$, the number of steps per policy roll out $k_{\text{step}}$, the mini-batch size $B$, the number of epochs $K$ to update the policy,  the discount rate $\gamma$,  the surrogate clipping coefficient  $\epsilon$, the entropy coefficient $c_2$,  and the value loss coefficient $c_1$. Besides, to stabilize training, we use the gradient clipping strategy and set the maximum gradient norm allowed to be $g_{\max}$. We also use the learning rate annealing technique, which aims to linearly decay the learning rate. The decay equation is given by
\begin{equation}\label{linear_decay}
    \alpha_k \leftarrow  \left(1-\frac{k-1}{M}\right) \times \alpha_0,
\end{equation}
where $\alpha_0$ is the initial learning rate. In reality, we find the algorithm performance is not sensitive to hyperparameters. All examples in this work share the same RL-specified hyperparameters. We present these hyperparameters in Table \ref{tab:RL_hyper}. 
\begin{table}[htbp]
    \centering
    \begin{tabular}{|c|c|} \hline 
         Hyperparameter name&  Value\\ \hline 
         $M$& 
    $50,000$\\ \hline 
 $B$&64 or 128\\ \hline
 $K$&4\\\hline
 $k_{\text{step}}$&512\\\hline
 $\gamma$&0.99\\\hline
 $\epsilon$&0.2\\\hline
 $c_1$&0.5\\\hline
 $c_2$&0.01\\\hline
 $g_{\max}$&0.5\\\hline
 $\alpha_0$&$4\times 10^{-4}$\\\hline
\end{tabular}
    \caption{The hyperparameters used in RGRL algorithm for all control examples. Fine-tuning the these hyperparameters may further improve the training efficiency.}
    \label{tab:RL_hyper}
\end{table}
The RL-specified hyerparameters can be further optimzied to enhance the performance such as the sample efficiency.

Moreover, the hyperparameters of actor and critic neural networks are presented in the following. All actor and critic neural network contain three layers. The first layer maps the partial observation (neural representation of current quantum state) into the next hidden layers. The number of input neurons equals to the dimension of the neural representation $d$. The hidden layer contains $64$ or $128$ neurons and the last output layer contains only single neuron for citric network and $n_{\text{actions}}$ neurons for actor network. The activation function used in RGRL is $\text{Tanh}(\cdot)$ non-linear function. Note that in case $d=32$, the number of neurons in hidden layer is set to be $64$. In case $d=96$, the number of neurons in hidden layer is set to be $128$. 

For different examples, the dimension of neural representation and the number of actions are also various. In all examples, we make use of discrete policy to control the parameters of quantum systems. Generally, to enhance the control efficiency, we make use of multi-discrete control policy to interact with the environment except for the XXZ model. It is quite intuitive since the parameters of quantum systems or Hamiltonian can be viewed independently from each other. For example, for Ising example, the number of actions is $18$, i.e. $n_{\text{actions}}=18$ as each parameter $J_i$ has three possible actions, namely $+1, -1$ and $0$. For XXZ model, the actions is set to be $8$ as there are $n_{\text{actions}}=8$ possible movements in the 2-dimensional parameter space except for the case of no movement. More specifically, each parameter can move towards left or right. Only one parameter moves or two parameters move simultaneously. For Kerr system, it is desirable to control the amplitude and phase of the light field. The amplitude and phase can be independently controlled and each has $3$ possible movements. Therefore, there are $n_{\text{actions}}=6$ possible movements in this environment. For quantum state retrodiction, the number of actions are also set to be $n_{\text{actions}}=6$ as the gate parameters of initial quantum state control the amplitude and phase of one coherent state. 

One more important hyperparameter is the discrete action step $\Delta a$ for each control example.  The specific value is presented in Table \ref{tab:actions}.
\begin{table*}[htbp]
    \centering
    \begin{tabular}{|c|c|c|} \hline 
         Quantum Model&  Discrete action step &Number of observales to measure\\ \hline 
         Random Ising model& 
    $\Delta a=0.05$ &$n_{\text{obs}}=5$\\ \hline 
 XXZ model& $\Delta a=1$ &$n_{\text{obs}}=50$\\ \hline 
 Kerr model& $\Delta a=0.3$ (linear decay)& $n_{\text{obs}}=3$\\ \hline 
 Generate output state of an unknown process& $\Delta a=(0.09, 0.06\pi )$&$n_{\text{obs}}=3$\\ \hline\end{tabular}
    \caption{The action steps and number of observables to measure for each quantum system.}
    \label{tab:actions}
\end{table*}
For Kerr model, the action step in each control loop is linearly decayed following the same rule with Eq.~(\ref{linear_decay}). It is worthy noting that the discrete action step will determine the convergence of the proposed algorithm. A moderate action step is necessary to obtain a higher fidelity as the fidelity may be sensitive to control parameters. During the late stage of the training, the action step should be small enough to ensure the convergence with high fidelity. In general, all examples can make use of the strategy of linear decay strategy. However, it is not our focus in this work to optimize this hyperparameters to obtain the optimal performance.

\section{Additional results on Control of disordered Ising model ground state}
We apply our RL algorithm for controlling the ground state of a disordered Ising model towards a target ground state.  Specifically, suppose the quantum state under control is a $6$-spin ground state of Hamiltonian 
\begin{equation}
\label{eq:Ising}
H_I:=-\sum_{i=0}^{4} J_i \sigma_i^z \sigma_{i+1}^z -\sum_{j=0}^{5} \sigma_j^x,
\end{equation}
where each $J_i\in (-1, 1)$ is an independent parameter. In this task, at each step, we must determine the change of each individual parameter, yielding a difficult high-dimensional control problem.

\begin{figure}[htbp]
    \centering
    \includegraphics[width=0.5\textwidth]{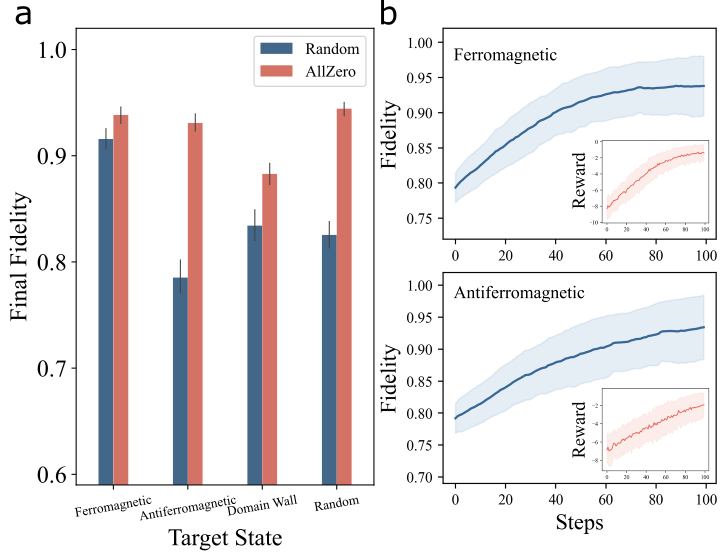}
    \caption{Control of Ising model ground states. Subfigure {\bf a} shows the quantum fidelity between the controlled state and the corresponding target state in the four different control scenarios after $100$ control steps averaged over $100$ experiments. The vertical line at the top of each bar represents the confidence interval with confidence level $95\%$. Subfigure {\bf b} shows the quantum fidelity of the controlled state in $100$ control steps, together with the reward at every step.}
    \label{fig:Ising}
\end{figure}

We suppose the quantum state under control is initialized either in the state $\ket{+}^{\otimes 6}$ that is the ground state when $J_i=0$ for each $i$, or in a ground state corresponding to a randomly parameterized model in Eq.~(\ref{eq:Ising}). We consider four control scenarios, each of which corresponds to a different target state: (1) a ground state in ferromagnetic phase corresponding to $J_i=0.8$ for $0\le i\le 5$, (2) a ground state in antiferromagnetic phase corresponding to $J_0=J_2=J_4=0.8$ and $J_1=J_3=J_5=-0.8$, (3) a ground state corresponding to $J_0=J_1=J_2=0.8$ and $J_3=J_4=J_5=-0.8$, and (4) a ground state corresponding to a randomly chosen Hamiltonian parameters.  We randomly choose five $6$-qubit Pauli bases out of $3^6$ possibilities and measure each single qubit, recording the outcome frequency distributions. At each control step, we perform the same set of quantum measurements. The measurement outcome statistics are fed into the neural network and then the RL algorithm outputs the actions on all six independent Hamiltonian parameters, where the action on each parameter is chosen out of the set $J_i \leftarrow \{J_i +0.1, J_i, J_i-0.1\}$.  

Figure~\ref{fig:Ising}{\bf a} illustrates the quantum fidelity between the controlled state after $50$ control steps and the target state in four different control scenarios including both types of quantum initial states. The quantum fidelity achieved starting from $\ket{0}^{\otimes 6}$ is higher than those achieved starting from a randomly disordered initial state, indicating that controlling a general disordered state towards a target state is much more difficult.  In Figure~\ref{fig:Ising}{\bf b}, we show both the quantum fidelity and reward for the control scenarios (2) and (3) at every control step from the beginning to $100$ steps. 

\section{Additional results on Control of phase transition in many-body systems}

\begin{figure}[htbp]
    \centering
     \includegraphics[width=0.45\textwidth]{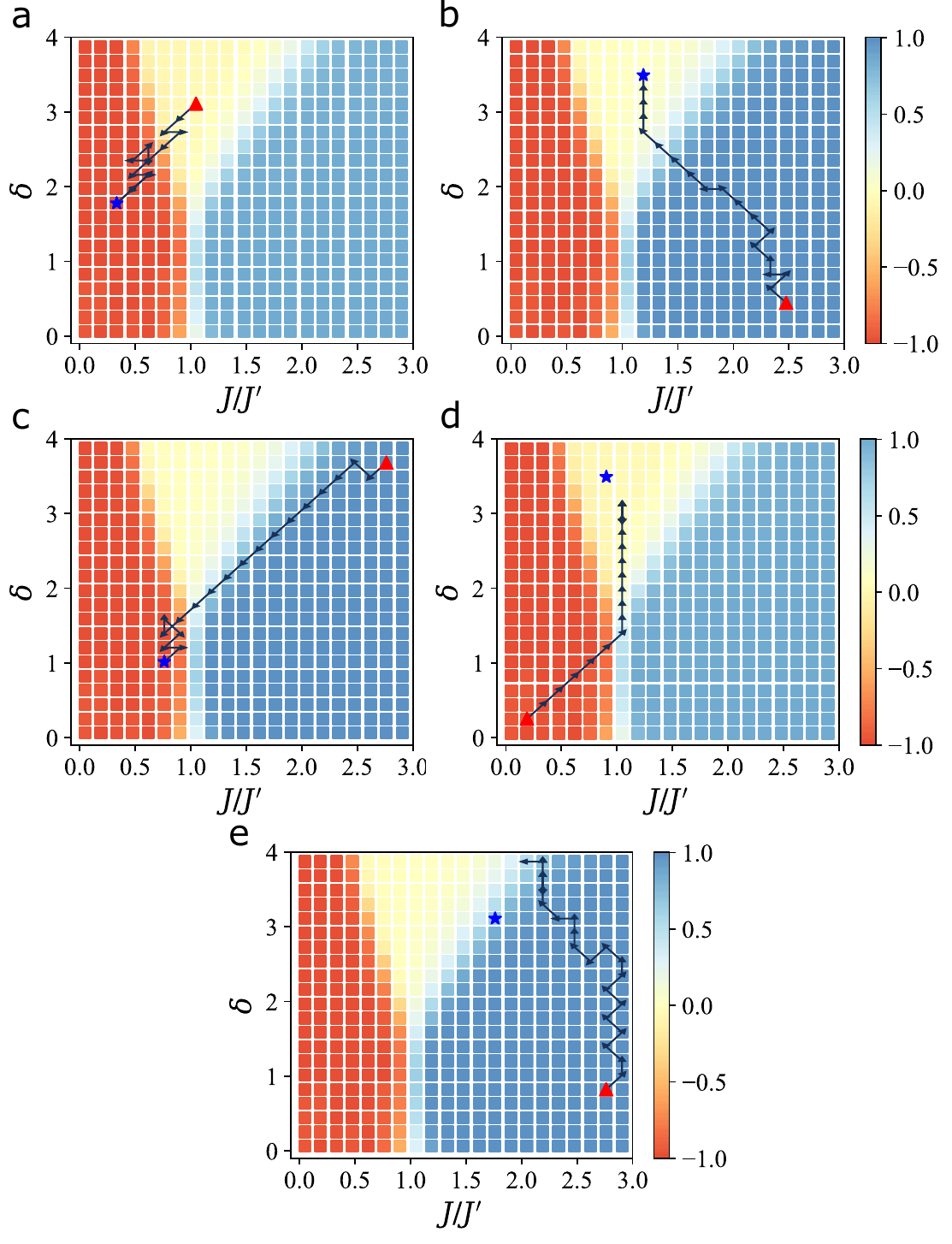}
    \caption{Trajectory of the controlled state, utilizing state representations to predict measurement statistics: {\bf a} from symmetry broken phase to topological SPT Phase, {\bf b} from trivial SPT phase to symmetry broken phase, {\bf c} from trivial SPT phase to topological SPT Phase, {\bf d} from topological SPT Phase to symmetry broken phase, {\bf e} from the trivial SPT phase to the phase boundary in the phase diagram.}
    \label{fig:XXZ_nolabel_trajec}
\end{figure}

\begin{figure}[htbp]
    \centering
     \includegraphics[width=0.45\textwidth]{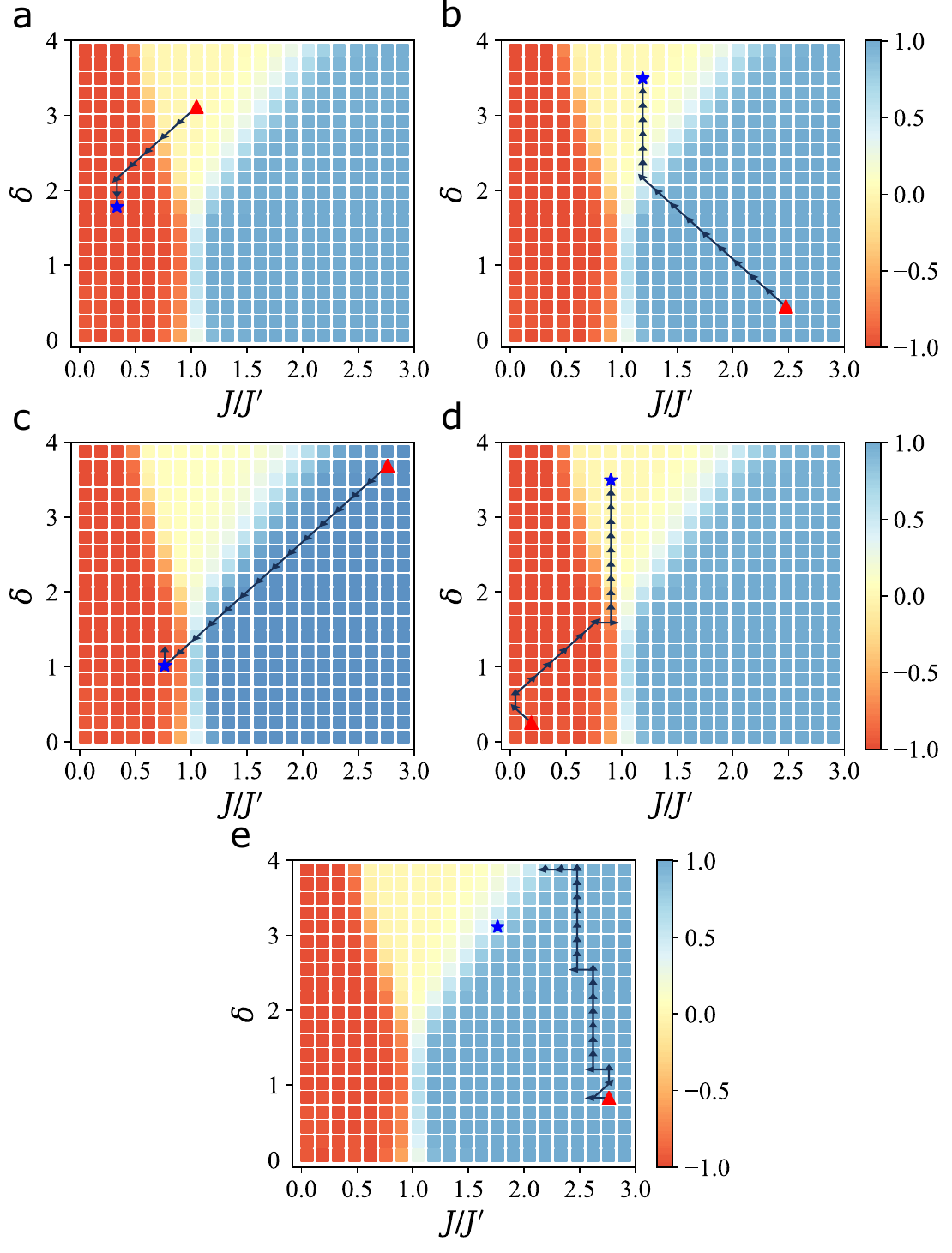}
    \caption{Trajectory of the controlled state, utilizing state representations to predict mutual information: {\bf a} from symmetry broken phase to topological SPT Phase, {\bf b} from trivial SPT phase to symmetry broken phase, {\bf c} from trivial SPT phase to topological SPT Phase, {\bf d} from topological SPT Phase to symmetry broken phase, {\bf e} from the trivial SPT phase to the phase boundary in the phase diagram.}
    \label{fig:XXZ_label_trajec}
\end{figure}

In this section, we present additional examples of controlling many-body ground states near phase transitions in the bond-alternating XXZ model. Figure~\ref{fig:XXZ_nolabel_trajec} illustrates the trajectories of ground state evolution under control, utilizing state representations to predict measurement statistics. In contrast, Figure~\ref{fig:XXZ_label_trajec} presents the trajectories based on state representations to predict mutual information.

\begin{figure*}[htbp]
    \centering
     \includegraphics[width=\textwidth]{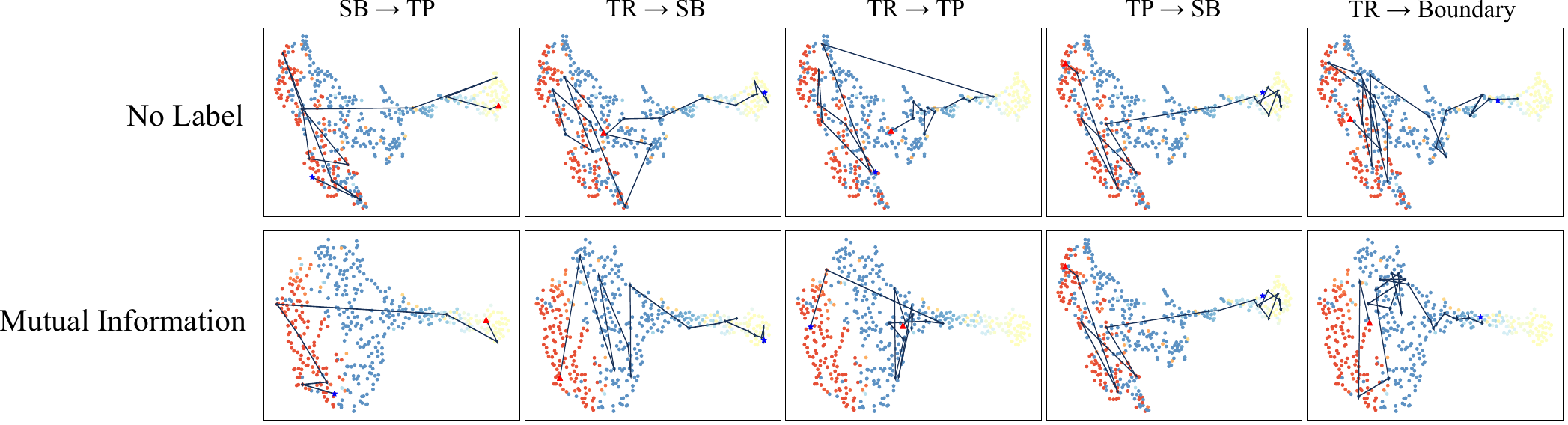}
    \caption{Trajectories of the
    controlled state under control in the representation spaces.}
    \label{fig:XXZ_embed_trajec}
\end{figure*}

As demonstrated by the results, the RGRL algorithm performs better when using state representations trained to predict mutual information compared to those trained merely to predict measurement statistics. This can be explained by the fact that the representation network leverages the nonlinear properties of the states, allowing the state representations to retain more comprehensive information. Consequently, this enhanced informational richness in the state representations likely contributes to the improved performance of the RGRL algorithm in controlling many-body ground states. We present the trajectories of the controlled states in the representation space in Figure~\ref{fig:XXZ_embed_trajec}. It can be observed that the positions of state representations are more separate when mutual information is used, enabling the algorithm to construct the correct trajectory with fewer steps.

\section{Additional results on Control of preparing continuous-variable cat states}

\begin{figure}[htbp]
    \centering
     \includegraphics[width=0.3\textwidth]{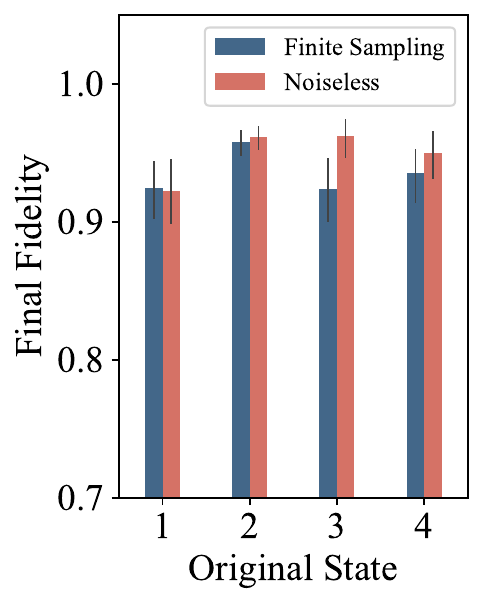}
    \caption{Control of preparing cat states in the scenario where measurement quadratures are randomly selected at each control step. The figure shows the quantum fidelity between the controlled state and the target cat state in four different scenarios after 20 control steps, averaged over 100 experiments.}
    \label{fig:CV_performance_rand}
\end{figure}

In this section, we demonstrate a different scenario in which we randomly select three different measurement quadratures at each control step, rather than keeping them consistent throughout the entire process of preparing the target cat state. We utilize the same four scenarios to test the performance of our proposed algorithm. Figure~\ref{fig:CV_performance_rand} shows the quantum fidelity between the controlled state and the target cat state in these four different scenarios after 20 control steps.

As shown in the results, the performance of our algorithm in this scenario surpasses its performance in the scenario where the measurement quadratures are fixed throughout the entire process. We believe this improvement is because the neural network collects more comprehensive information about the state when the measurement quadratures are varied. This variability likely enables the network to capture a richer set of state features, thereby enhancing the control precision and ultimately leading to higher quantum fidelity.

\bibliography{supp.bib}